\documentclass[onecolumn,preprintnumbers,amsmath,amssymb,prl]{revtex4}
\usepackage{pifont}
\usepackage{amsmath}
\usepackage{amsfonts}
\usepackage{graphicx}
\usepackage{amssymb}
\usepackage{dcolumn}
\usepackage{bm}
\usepackage{slashed}

\begin{document}

\title{Seed and vacuum pair production in strong laser field}

\author{Huayu Hu}
 \email{huayu.hu.cardc@gmail.com}
 \affiliation{Hypervelocity Aerodynamics Institute, China Aerodynamics Research and Development Center, Mianyang 621000, Sichuan, People's Republic of China}

\date{\today}

\begin{abstract}
Researches on the electron-positron pair production in the presence of the intense laser field are reviewed, motivated by the theoretical importance of the nonperturbative QED problem and the worldwide development of the strong laser facilities.  According to distinct experimental requirements and theoretical methods, two types of pair production are elaborated, which are  respectively the pair production in the combination of a seed particle and the strong laser, and vacuum pair production without a seed particle. The origin of the nonperturbative problem caused by the strong field is analyzed. The main ideas, realization, achievements, validity, challenges and bottleneck problems of the nonperturbative methods developed for each type of the pair production problem are discussed.
\end{abstract}


\maketitle

\section{I. Strong field QED and strong laser facilities} 
Quantum electrodynamics (QED) is the fundamental theory for the light-particle interaction. The theory has passed various tests of scattering experiments, and also remarkable agreements are achieved on radiative correction measurements, for example the best experimental value and the best theoretical value of the anomalous magnetic moment of the electron are within agreement of $10^{-12}$ \cite{anorev}, the accuracy of which is `equivalent of measuring the distance from the Earth to the Moon to within the width of a single human hair' \cite{anorev1}. Conventional study of QED processes assumes no or at most a weak background field, where the theoretical calculation can be carried out in a perturbative way according to the small coupling constant $\alpha\simeq 1/137$ between the electromagnetic field and the particle. In comparison, processes in a strong electromagnetic field have not been studied thoroughly.

One representative QED process in a strong field is the fermion pair production, that a strong field can excite the ephemeral virtual particles fluctuating in the vacuum into real states, while weak field problems in contrast always assume a stable constant vacuum as the ground state. At the dawn of QED in 1930s Sauter and Schwinger predicted that a static electric field $E$ can produce real electron positron pairs, which are the lightest and thus the most possible to be produced charged particles, at the rate of $R\propto \exp(-\pi E_c/E)$ where the critical value is $E_c\approx 1.3\times10^{16}\,$V/cm and the corresponding field intensity is $I_c\approx 2.3\times10^{29}\,$W/cm$^2$ \cite{sauter, schwinger}. The amount of work exerted on an electron by the critical field $E_c$ over a  Compton wavelength is $m=0.511$MeV which is the electron/positron mass (rest energy). When $E\geq E_c$ the vacuum boils up with electrons and positrons popping out. A strong electric field can be found in the vicinity of a heavy nucleus. It has been pointed out that near a nucleus of a point charge $Z$ with $Z\sim1/\alpha$ the $1s$ state dives into the Dirac sea and electron-positron pair is produced, and a point charge of $Z>1/\alpha$ can not exist due to the pair production while extended higher charges can in principle exist\cite{GreinerStrong}. Lacking the existence of stable nuclei with such high charges, it is proposed to use the strong field in the collision of two very heavy ions, especially when the two ions form a quasi-molecule. However, for the pair production in the relativistic heavy ion collision the leading-order contribution is the perturbative two-virtual-photon mechanism \cite{bremCollisionPP}. Besides, when projectile energies are above 6 MeV/nucleon violent nuclear reactions set in that could provide a formidable background in positron spectra \cite{GreinerStrong}.


Recent years have witnessed the fast development of the strong laser technology based on the Nobel prize invention of the chirped pulse amplification.
High field intensity is achieved by suppressing the laser energy into a short duration and focusing into a small region. As the laser intensity $10^{11}$W/cm$^2$$\sim10^{18}$W/cm$^2$ is achieved which is comparable to or stronger than that of the Coulomb field experienced by the bound electrons, abundant interesting phenomena are discovered in the interaction between the laser and atoms, molecules and plasmas, such as the above threshold ionization, high harmonic generation, laser wake field acceleration and so on. As the laser technology provides a much more precise and controllable way to produce a strong field, high agreement has been achieved between the experiment and the theory, and the established good understanding of the processes promotes novel applications, such as the attosecond light source, table-top particle accelerator, and so on. Laser intensities as high as $\sim 10^{22}$W/cm$^2$ have been claimed by focusing a 300TW (1TW=$10^{12}$W) laser (upgrade of the HERCULES laser\cite{HERCULES}) into a diffraction-limited 1.3$\mu$m focal spot\cite{Yanovsky2008}. This year, ELI-NP (Romania)\cite{ELINP} reported the accomplishment of two 10 PW (1PW=$10^{15}$W) laser arms able to reach intensities of $10^{23}\,$W/cm$^2$ and electrical fields of $10^{15}\,$V/m, and the production of brilliant $\gamma$ beam with photon energy up to 19.5 MeV, obtained by incoherent Compton back scattering of the intense laser on an electron beam. SULF (China) has also realized 10PW laser amplification \cite{SULF1}. The $\sim10$PW lasers under development include also ELI Beamlines (Czech Republic) \cite{ELIbeam}, Apollon (France) \cite{Apollon}, Vulcan (UK) \cite{Vulcan}, and so on. Laser intensities of the order of $10^{24}$W/cm$^2$ or even higher are promising in the near future. Moreover, Exawatt (1 Exawatt=$10^{18}$W) laser is planned to be constructed at SEL (100PW, China) \cite{SEL} in the next five years, and also at ELI (200PW) \cite{ELI} and XCELS (200PW, Russia) \cite{XCELS}, aiming to achieve the intensity $\geq10^{26}$W/cm$^2$. In these high intensity fields, particle dynamics is dominantly relativistic and notably influenced by its own radiation (radiation reaction), the index of the refraction of the vacuum can be changed (vacuum polarization),  energetic gamma photons can be produced via nonlinear Compton scattering or laser-assisted bremsstrahlung, and so on.

Although at present and in the foreseeable future the available laser intensity is still several orders lower than that of the critical electric field $I_c\approx10^{29}$W/cm$^2$, laser induced electron-positron pair production has already been achieved experimentally. The SLAC-144 experiment in 1996 using a 46.6GeV electron beam colliding with a $I\sim 10^{18}$W/cm$^2$ and $\omega=2.4$eV laser realized the multi-photon pair production \cite{SLAC}. Relativistic quasi-monoenergetic positron jets were produced by shooting short ($\sim 1$ps) intense laser pules $I\sim 10^{20}$W/cm$^2$ at thick gold targets via Bethe-Heitler mechanism \cite{chen1,chen2}. The common feature of these experiments is that an extra high energy particle/photon works together with the laser to produce the pair. The energetic particle/photon can be produced by an external accelerator, or by the laser field itself through direct acceleration of charged particles, nonlinear Compton scattering, and so on. The characteristics of these processes depend strongly on the parameters of the laser field and the particle/photon, and there are three Lorentz-invariant dimensionless parameters widely used in the study of the pair production as well as other strong field QED processes. One is the multi-photon classical nonlinearity parameter $\xi=e\sqrt{-\langle A_{\mu}A^{\mu}\rangle}/m=e {\bar E}/(m\omega)$ in the natural units $\hbar=c=1$ and the Heaviside-Lorentz units with the electron absolute charge $e=\sqrt{4\pi\alpha}$, where $A$ is the four-vector potential of the laser field,  $\bar E$ is the root-mean-square value of the electric field component, $\omega$ is the laser photon energy, and $\langle \,\rangle$ represents time averaging. $\xi$ has simple explanations as the ratio of the work done by the field on an electron over a wavelength to the electron mass, and as the ratio of the work done over the Compton wavelength to the photon energy. Therefore, when $\xi\gtrsim 1$ the probability of multi-photon absorption becomes notable and the dependence on the field becomes nonlinear \cite{Ritus}. Besides, since the momentum of a plane-wave-laser-driven-electron quiver oscillation can be estimated as $q_\approx eE/\omega=\xi mc$, $\xi>1$ indicates the relativistic particle dynamics \cite{Fedotov2016}. A quick calculation can be performed as $\xi\approx 6\times 10^{-10}\lambda[\mu$m]$\sqrt{I[\text{W/cm}^2]}$, and thus an optical laser with the wavelength $\lambda\approx1\mu$m and intensity $I\approx 10^{23}$W/cm$^2$ has $\xi\approx 10^2$. The quantum nonlinearity parameter is defined as $\chi=e\sqrt{-\langle(F_{\mu\nu}p_{\nu})^2\rangle}/m^3=\gamma(1+\beta)\bar E/E_{c}=\xi \omega_r/m$, where $p_\nu$ is the momentum of the electron colliding with the field, $F_{\mu\nu}$ is the field-strength tensor, $E_c$ is the critical field given above, and $\omega_r=\gamma(1+\beta)\omega$ is the photon energy in the rest frame of the electron. The last two expressions are derived assuming the electron collides with the laser head-on with $\gamma=p^0/m$ and $\beta=|\mathbf{p}|/p^0$. The parameter $\chi$ can be understood as the ratio of the energy that a particle absorbs from the laser field along a Compton wavelength over the electron rest energy, or as the ratio of the electric field seen by the particle in its rest frame to the critical value. When $\chi\sim 1$, the electric field strength in the electron rest frame is of the order of the critical value $E_c$.  For the available laser with $I\approx10^{23}$W/cm$^2$ and $\lambda\approx1\mu$m colliding head on with a 1$\,$GeV electron which can be obtained by laser wake field acceleration, $\chi\approx 1$. Quantum recoil becomes significant as $\chi\gtrsim 1$ and thus in this regime it is necessary to take into account quantum processes in calculating particle dynamics. It marks the quantum regime of laser-matter interaction. Classification of physical regimes of laser-matter interactions according to these parameters is reviewed in \cite{Fedotov2016}. If the colliding particle is a photon with momentum $k'$, the quantum nonlinearity parameter is
$\kappa=e\sqrt{-\langle(F_{\mu\nu}k'_{\nu})^2\rangle}/m^3=2\omega'\bar E/(m E_{cr})$, with $\omega'=k'^0$ being the colliding photon's energy. The last equality is again valid when a head-on collision is assumed.

It is worthy noting that producing the matter from the light is not mysterious according to the famous Einstein's formula  $E=mc^2$. For example, there is certain probability for two head-on colliding gamma photons of the energy above $m=0.511$MeV to produce an electron-positron pair. Such processes $\gamma+\gamma'\rightarrow e^-+e^+$ are called the Breit-Wheeler process \cite{BW1934}, and the cross-section can be analytically obtained by perturbative Feynman rules. But in the strong laser, numerous photons act coherently and interact nonperturbatively with the particle, significantly changing the behavior of the particle's dynamics, and particle's radiation and absorption. Besides real particles, the strong field can drive the virtual particles so notably that the property of the vacuum can be changed such as vacuum birefringence/dichrotism and also the structure of the vacuum under disturbance can be detected, see the review \cite{BenVPReview}. Much effort has been devoted to develop the nonperturbative theoretical and numerical methods for the strong field QED problems. In this paper, we mainly review the theoretical research development on the electron-positron pair production in a strong laser. Two types of pair production with distinct experimental requirements and theoretical approaches are considered: (1) pair production with a seed particle such as in the case of SLAC E-144 experiment; and (2) vacuum pair production without a seed particle. We discuss problems that have been successfully treated along with the non-perturbative methods developed correspondingly, and problems that remain challenging.

\section{II. Pair production in the laser field with a seed particle}
\begin{figure}[h]\centering
\includegraphics[height=6cm,width=10cm]{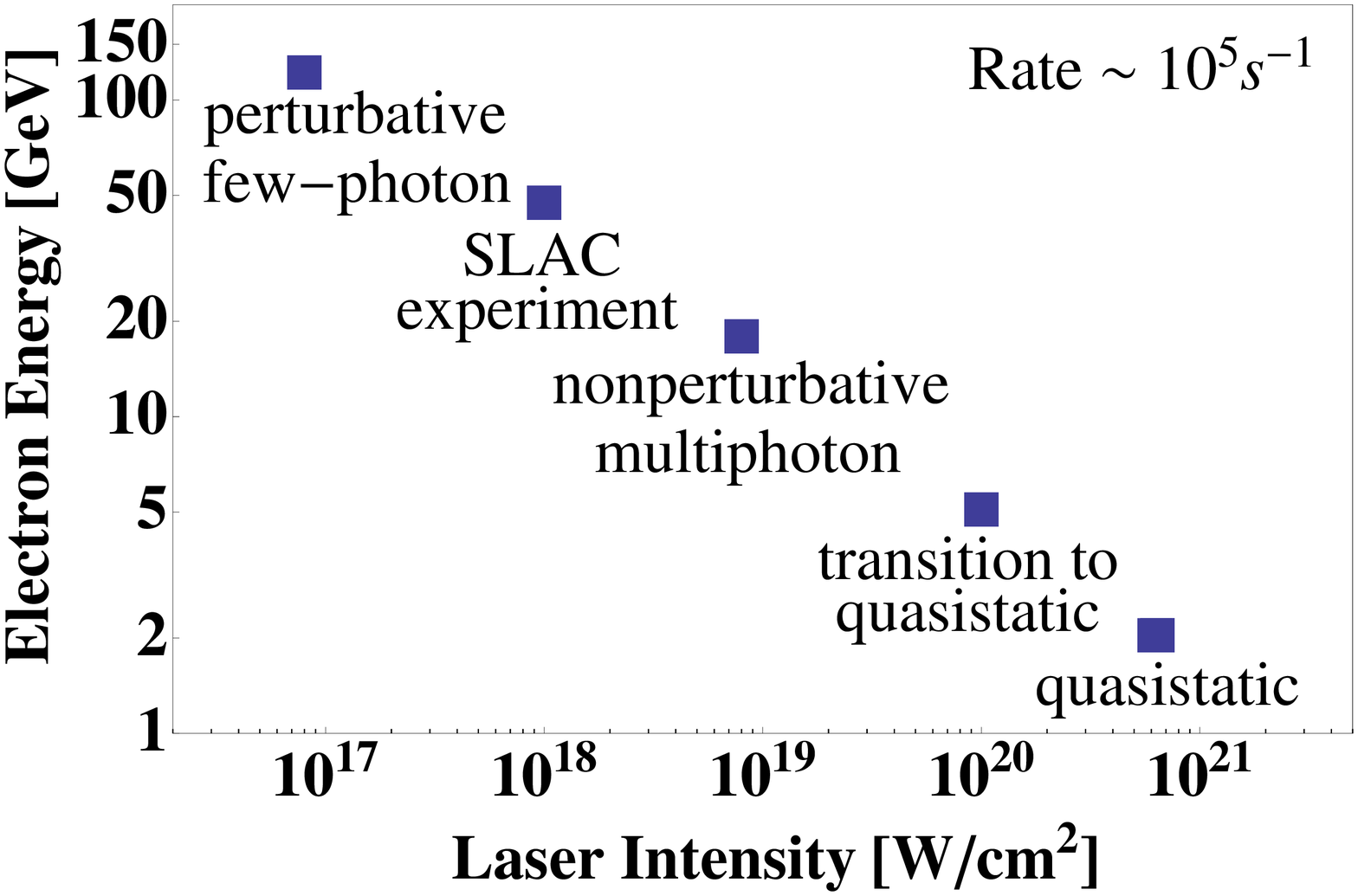}
\vspace{-0.3cm}
\caption{\label{ratefigure} SLAC-E144 experiment \cite{SLAC} achieves a $e^+e^-$ pair production rate $\sim10^{5}$s$^{-1}$ in electron-laser collisions. Various combinations of the electron energy and the laser intensity with the fixed laser photon energy 2.4eV can give the same production rate. The parameters transits from the perturbative to the fully nonperturbative regimes. Adapted from \cite{Hu2010}.}
\end{figure}

Not long after the invention of the laser in 1960s, people began to consider the possibility of using it to produce the electron-positron pair via multi-photon absorption, and found a detectable number of pairs can not be obtained as the contemporary laser field was far weaker than the critical field \cite{1970pairlaser1,1970pairlaser2}. Since this is still the case at present and in the foreseeable future, a seed particle is vital to realize the detectable pair production as demonstrated by the experiments \cite{SLAC, chen1, chen2}. Generally speaking, for the setup of a seed particle colliding with the laser field, the higher the energy of the particle is, the lower the intensity of the laser is required to achieve the same pair production rate, as shown in Fig. \ref{ratefigure}. It can be understood as the laser intensity (in fact as well as the frequency) gets enhanced in the rest frame of the seed particle. Indeed in the SLAC E-144 experiment, the colliding electron sees in its rest frame the laser field with the intensity close to $I_c$. However, it is not appropriate to claim that the pair production in the critical field regime is simulated, because the governing parameters $\xi$, $\chi$ and $\kappa$ are Lorentz invariant.

\begin{figure}[t]\centering
\includegraphics[height=6cm,width=10cm]{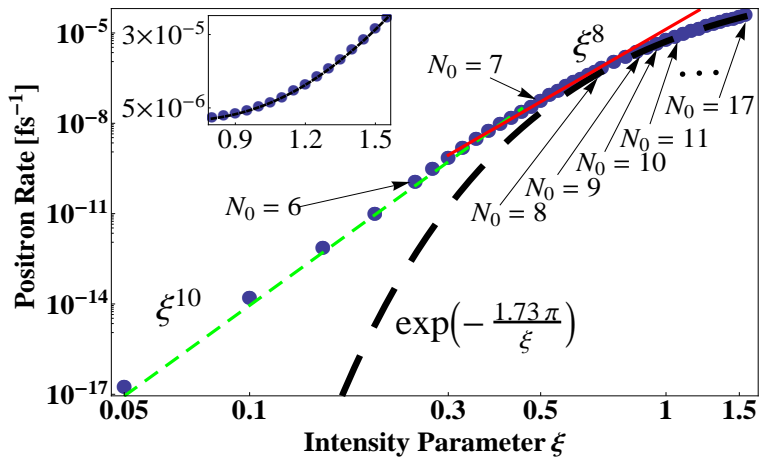}
\vspace{-0.3cm}
\caption{\label{power2} Dependence of $e^+e^-$ production rate (blue dots) on $\xi$ in the electron-laser collision with the same initial electron energy 46.6\,GeV and laser photon energy 2.4eV as in the SLAC E-144 experiment \cite{SLAC}. The strong-field features are shown as (1) Kinematically a 46.6\,GeV electron absorbing five 2.4eV photons can produce a $e^+e^-$ pair in the vacuum. But with the heavier effective mass $m^*=m \sqrt{1 + \xi^2}$ increasing with $\xi$, more photons must be absorbed to reach the threshold of pair production. The numbers $N_0=6,\ldots,17$ indicate the minimum numbers of laser photons that must be absorbed for the generation of a $e^+e^-$ pair in different $\xi$ regimes. (2) The power law fits $\xi^{10}$ (green short-dashed line) and $\xi^{8}$ (red solid line) which respectively would correspond to 5 and 4 photons absorption in a perturbation theory actually take place where at least 6 (see $\xi\gtrsim 0.3$) and 7 (see $\xi\gtrsim0.5$) photons must participate. (3) The exponential fit $\exp[-1.73\pi/\xi]$ as $\xi\gtrsim 1$ resembling the Sauter-Schwinger pair production rate in a constant electric field manifests a clear nonperturbative feature. Adapted from \cite{Hu2010}.}
\end{figure}

Crudely speaking, in case the seed particle is an energetic photon the pair production can be seen as a generalization of the Breit-Wheeler (BW) mechanism, which is kinematically allowed as long as the two photon momentum $k_1$ and $k_2$ satisfies $k_1\cdot k_2\geq 2m^2$. In case the seed particle is a heavy ion, the pair production can be seen as a generalization of the Bethe-Heitler (BH) mechanism $Z_{\infty}+\gamma\rightarrow Z_\infty+e^++e^-$ \cite{BH1934}, which is the conversion of a photon into an electron-positron pair near an infinitely heavy nucleus and thus photon energy should satisfy $E_\gamma\geq 2m$. In case the seed particle is not so heavy such as an electron, the pair production can be seen as a generalization of the trident process $e^-+\gamma\rightarrow e^++2e^-$ where $\gamma$ can either be a real photon \cite{trident} or a virtual photon of the Coulomb field \cite{1974pairlaser}. The BW, BH and trident processes can be calculated by perturbative QED methods such as the Feynman rules. Due to the small coupling constant $\alpha\sim 1/137$, the cross-section of the processes with $n$-photon absorption is suppressed as $\alpha^n$. However, in the perturbation theory the rate of a $n$-photon absorption process should also be multiplied by parameters proportional to $I^n$ with $I$ being the photon field intensity. It can be speculated that the trend of the decreasing pair production contribution with the increasing number of photon absorption would be changed if the laser intensity is very high, and therefore many channels of multiple photon absorption must be taken into account. It can be seen also from the point of view of QED with a classical field $A_\mu$. Since the interaction term in the Hamiltonian is $eA_{\mu} \hat{J}^\mu$ where $\hat{J}=\hat{\bar{\Psi}}\gamma^\mu\hat{\Psi}$ is the quantized particle current, the expansion parameter of the QED is $\propto eA/m=\xi$, and the validity of perturbation expansion is in question as $\xi\geq 1$ \cite{Fedotov2016}. Besides, there is another serious theoretical difficulty as the effective coupling `constant' can change from order to order in an intense electromagnetic field. This is because the particle current $\hat{J}(t)$ is also notably influenced by the field as a collective effect of numerous laser photons working coherently on the particle. For clarity we can take a look at the analytical solution of the Dirac equation $(\gamma_\mu (\hat{p}^\mu-eA^\mu)-m)\Psi=0$ in a classical plane-wave field $A^\mu(k\cdot x)$ with $k^2=0$, known as the Volkov state \cite{Volkov, BrownKibble}. In Lorentz gauge $k\cdot A=0$, an electron wavefunction can be solved as
\begin{equation}\label{Volkovs}
 \Psi_{p,s}=\sqrt{\frac{m_e}{q^0 V}}(1-\frac{e\slashed{k}\slashed{A}}{2k\cdot p})u_{p,s}e^{i f}\,,
\end{equation}
where $V$ is the normalization volume, $q^0$ is the time-averaged energy of the particle in the field, $u_{p,s}$ is the free particle spinor initially outside the field with the free momentum $p$ and spin $s$, and for a linearly polarized field $A^\mu=a \epsilon^\mu \cos(k\cdot x)$ with the amplitude $a$ and the polarization vector $\epsilon$ the phase term is
\begin{equation}\label{Volkovs plinear}
f=-q\cdot x+\frac{ea \epsilon\cdot p}{k\cdot p}\sin(k\cdot x)-\frac{e^2a^2}{8 k\cdot p}\sin(2 k\cdot x)\,,
\end{equation}
with $q=p+\frac{e^2a^2}{4 k\cdot p}k$. $q$ can be seen as a time-averaged dressed momentum, and the dynamics is featured by an effective mass $m^{*2}=q^2=m^2 (1 + \xi^2)$. When $\xi\gg1$, this can remarkably change the kinematic relations obtained for the free electron mass $m$, see Fig. \ref{power2}. Fourier expansion of the wavefunction shows \cite{BrownKibble,Ritus}
\begin{align}\label{VolkovBes}
 \Psi_{p,s}=\sqrt{\frac{m_e}{q^0 V}}u_{p,s}e^{-i q\cdot x}\sum_{-\infty}^{\infty}C_n(\alpha,\beta)e^{i n k\cdot x}\,,
\end{align}
where $C_n(\alpha,\beta)=J_n(\alpha,\beta)-\frac{e a \slashed{k}\slashed{\epsilon}}{4k\cdot p}(J_{n-1}(\alpha,\beta)+J_{n+1}(\alpha,\beta))$ with $\alpha=\frac{ea \epsilon\cdot p}{k\cdot p}$ and $\beta=\frac{e^2a^2}{8 k\cdot p}$. The generalized Bessel function $J_n(\alpha,\beta)$ is defined by the Fourier transform
\begin{displaymath}
J_n(\alpha,\beta)=\frac{1}{2\pi}\int_{-\pi}^{\pi}\exp[-in\theta+i \alpha \sin(\theta)-i \beta \sin(2\theta)]d\theta\,.
\end{displaymath}
A similar expansion like Eq. (\ref{VolkovBes}) can be obtained for the case of a circularly polarized plane wave laser, where the Fourier coefficients are ordinary Bessel functions, making it relatively simpler in the calculation.

According to Eq. (\ref{VolkovBes}), the Volkov state is the superposition of infinite states each with a well-defined four-momentum $q + n k$, $n=(-\infty,\infty)$. It can be seen as occupying a ladder of uniformly spaced energy levels, with the energy difference between subsequent levels being the energy of the laser photon  \cite{Eden}. This picture has been used in explaining the high-order harmonic radiation in laser-atom interaction as a result of the Coulomb field induced transition between these energy levels  \cite{Guo, Hu2008}. Intuitively, the coefficient $C_n$ indicates to some extent the coupling strength between the particle and $|n|$ laser photons. Due to the Bessel functions' property, in the weak field limit $\xi\rightarrow 0$ and thus $\alpha,\beta\rightarrow 0$ there is $C_n\propto \xi^{|n|}\sim e^{|n|} I^{|n|/2}$, which is the multiplication of the expansion coefficient $e^{|n|}$ of the perturbative Feynman diagram with $|n|$ photon lines and the intensity parameter $I^{|n|/2}$ of $|n|$ photon interaction. However, for a large $\xi$, $C_n$ does not follow the simple power law. This hints the rise of the nonperturbative character and breakdown of the perturbation method. The breakdown of the power law relation of $\xi$ as $\xi\rightarrow 1$ can also be observed in deriving the particle-propagator and the scattering matrix, and can also be directly seen in the curve of the pair production rates {\it e.g.} in Fig. \ref{power2}.

Some ideas for treating the nonperturbative QED problems in a strong field have been developed. First of all, the strong field is almost always approximated as an external classical field. This is due to the negligible change in the number of photons via absorption, radiation and scattering compared to the huge number of photons in the relevant field modes. For example, the number of photons of the laser with wavelength $\lambda=1\mu$m and intensity $I=10^{18}$W/cm$^2$ in the diffraction-limited focusing center (volume $V\approx\lambda^3$) reaches $N=IV/(c\hbar\omega)\approx 10^{-4}I$[W/cm$^2$]$\approx 10^{14}$ and thus the energy taken from the field to produce a pair is relatively negligible. Through calculating the particle propagator in a fixed classical electromagnetic field instead of the vacuum and with the knowledge of the single particle's eigen-energies of the quantum mechanical equation in a constant electric field, the pair production rate in a constant electric field is obtained nonperturbatively in the sense that in the tree level the interaction with the external field to all orders has been taken into account \cite{schwinger}. Similarly, with the knowledge of the Volkov state, the particle propagator as well as other QED quantities in a plane wave field can be obtained. The treatment can be formulated as the Furry picture \cite{Furry}. In this formalism, the free particle's Hamiltonian combined with the coupling term of the strong external field is taken as a new effective free Hamiltonian, the eigenstates of which are obtained by solving the corresponding quantum dynamical equations, like the Volkov states. The remaining interaction Hamiltonian includes weak perturbations between the dressed particle and other quantized photon modes only. Therefore, perturbative expansion of the evolution operator can be carried out with respect to the weak interaction, and Furry-Feynman rules/diagrams can be set up. Compared to the conventional Feynman rule/diagram, it replaces the free particle state to the new eigenstate (e.g., Volkov state), and the particle propagator in vacuum to the particle propagator in the strong field (e.g., Dirac-Volkov propagator).  In this way, the non-perturbative interaction between the particle and the strong classical field is accounted to all orders. Furry picture is a general treatment, and also applies successfully in strong field atom/molecular physics \cite{Eden, Guo, Hu2008}.

\begin{figure}[t]\centering
\includegraphics[height=4cm,width=10cm]{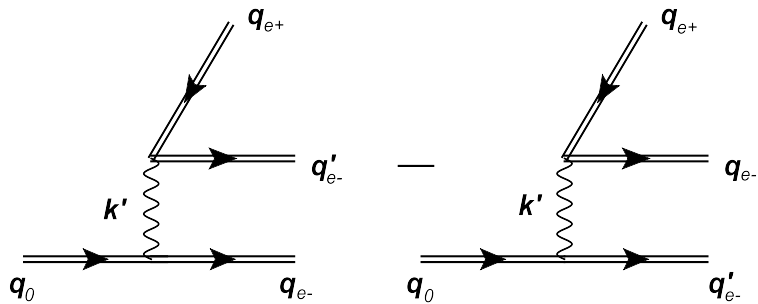}
\vspace{-0.3cm}
\caption{\label{trident} Furry-Feynman diagrams of multi-photon trident pair production in the electron laser
collision as in the SLAC-E144 experiment. The electron/positron states are Volkov states and are labeled by the laser-dressed particle momenta.}
\end{figure}

\begin{figure}[t]\centering
\includegraphics[height=6cm,width=10cm]{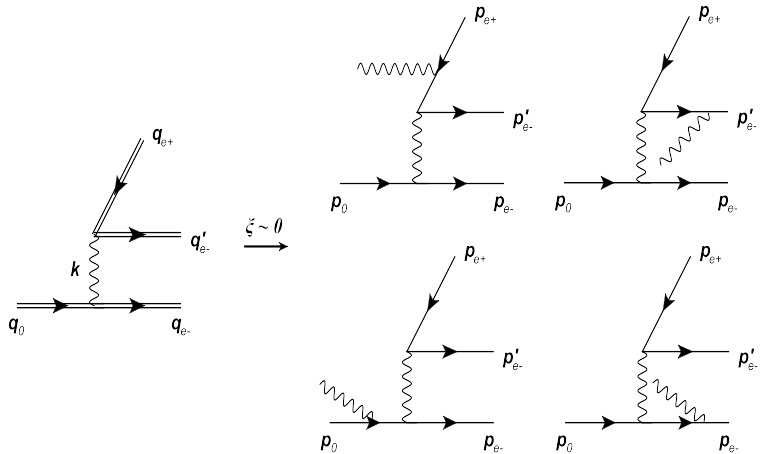}
\vspace{-0.3cm}
\caption{\label{tridentweak} In the weak field limit $\xi\rightarrow 0$, the Furry-Feynman diagrams of multi-photon trident pair production are reduced to the Feynman diagrams of trident pair production in the collision of one particle and one photon.}
\end{figure}

\begin{figure}[t]\centering
\includegraphics[height=4cm,width=12cm]{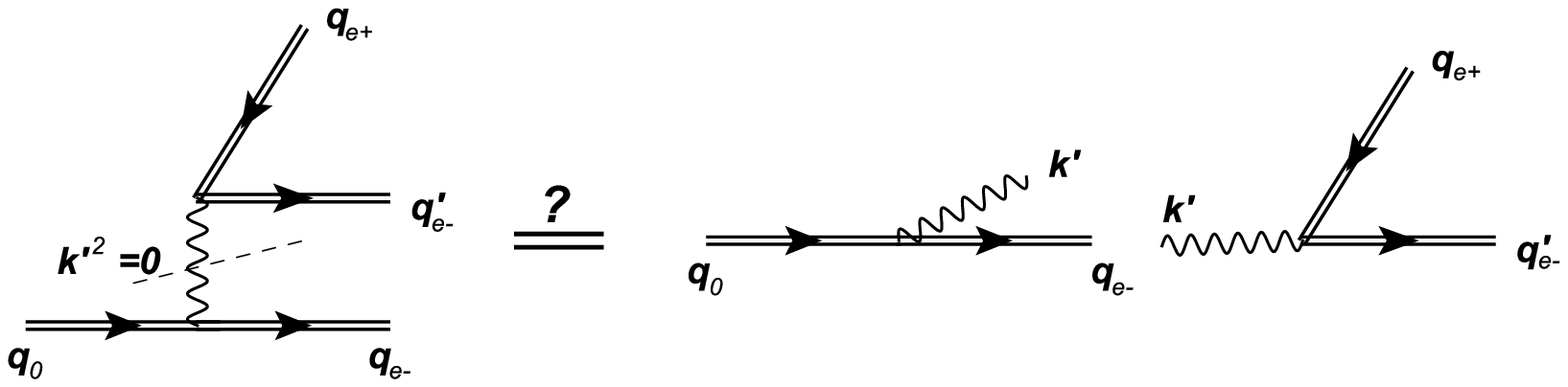}
\vspace{-0.3cm}
\caption{\label{tridenttwostep} When the field intensity is high, there are many possible channels of pair production with different numbers of photon absorption. In some channels it is possible for the intermediate photon to be on shell $k'^2=0$ and the photon propagator to diverge. The pole problem can be solved by a regularized photon propagator \cite{Hu2010, Huthesis}. The circumstance under which the multi-photon trident process with a pole in the photon propagator can be treated as a two-step process is also investigated \cite{piazzatrident, Dinutrident}.}
\end{figure}

Using the Furry-Feynman diagrams and the Volkov states, pair production of the BH type in relativistic nuclei colliding with intense plane-wave laser fields has been studied intensively. Total production rates have been calculated in various interaction regimes \cite{MVG,Avetissian,Sieczka,Milstein}. Energy spectra, angular distributions and spin correlations of the produced particles were obtained \cite{Sieczka,MVG,KuchievPRA,KaisaCorr, Carstenspin}. The influences of bound atomic states \cite{bf}, an additional high-frequency photon \cite{ADP}, and the nuclear recoil \cite{Sarah,KaisaRecoil} have been studied as well, see also the review \cite{Piazzareview}. In particular, the SLAC E-144 experiment is of the trident type, where the recoil of the electron as well as the indistinguishability of the two final electrons adds complexity to the derivation and computation. Figure \ref{trident} shows the Furry-Feynman diagrams of the process, and their lowest-order expansion in the weak field limit gives the Feynman diagrams of the trident process shown in Fig. \ref{tridentweak}. A more technically subtlety problem is the divergence problem induced by the pole of the photon propagator when the intermediate photon gets on shell $k'^2=0$ as shown in Fig. \ref{tridenttwostep}. The whole process can be `roughly' seen as a two step process of a nonlinear Compton scattering followed by a multi-photon pair production. Generally speaking, the calculation of such a process would require the input of the duration of the interaction time, as indicated by the dimensionality analysis that $[R_{\text{Total Pair Production}}]=[R_{\text{Nonlinear Compton Scattering}}]\times [R_{\text{Multiphoton Pair Production}}]\times [T_{\text{Interation duration}}]$. We have derived and proved analytically \cite{Huthesis} that $T_{\text{Interation duration}}$ can be involved as a regulator to the photon propagator, and thus the one-step process ($k'^2\neq 0$) and the roughly-speaking two-step process or the divergent channel ($k'^2=0$) can be calculated in a unified way \cite{Huthesis,Hu2010}. Note that the regularized propagator does not change the gauge covariance of a scattering matrix. Note also that it is not always accurate to directly use the formula of a two-step process even if the intermediate photon is energetically on shell. In \cite{piazzatrident}, it shows that such a photon can still be a virtual photon so that it is a one-step process if the photon has a longitudinal polarization. Another way to deal with the pole is to take into account the actual finite interaction time in the particle's wavefunction. This is realizable when the laser field is a short plane-wave pulse which is composed of multiple plane waves with the wave-vectors in the same direction \cite{ildertonPRL, kampfertridentpulse, Dinutrident}. In \cite{kampfertridentpulse} the effects of the pulse duration is transparently quantified by resorting to a special pulse model and in \cite{Dinutrident} the regimes are elaborated where the correction to the two-step part resulted from the exchange term is important. However, in the SLAC E-144 experiment the interaction time is not determined by the relatively long pulse duration but by the flying time $\sim 40$fs of the particle in the laser field. Numerical calculations with the regularized propagator \cite{Hu2010, Huthesis} have satisfyingly reproduced the experimental measurements of the total production rate and the positron momentum spectra. Strong field effects are illustrated in Fig. \ref{power2}.

Seed pair production in a plane-wave laser field of different configurations is explored. Effects of the laser polarization, short duration, bichromatic configuration, delay of consecutive pulses, and spin effects on the pair production of both the BH and BW type are systematically studied \cite{carsteninterferenceBHBichromatic, carstenEnhanceTwocolor, carstenInterferenceBW, carstenSpinShort, carstenBWTwocolorDelay}. An experimental laser beam can be treated approximately as a plane wave if the focus spot size is greater than 10 times of the wavelength. In view of the non-plane-wave strong fields such as tightly focused lasers, colliding lasers and so on,  it is needed to go beyond the Volkov state obtained in a plane-wave field. Ultrarelativistic quasiclassical wavefunctions with classical electron action in an arbitrary electromagnetic background have been derived and applied in calculating first-order (in the Furry picture) strong field QED processes including BW type pair production \cite{PiazzaVolkov}. Ultrarelativistic quantum wavefunctions of the Klein-Gordon equation have also been derived  \cite{Hu2015, BenHu2016}. In \cite{Ritus, RitusCC} it is proposed that when $\chi\gg 1$ the processes in the strong laser at each given moment are essentially determined by the instantaneous local crossed ($B\perp E$, $B=E$) field at that moment and the probability depends mainly on $\chi$ alone, known as the local constant crossed field approximation (LCFA). Since the analytical solution of the Dirac equation in a constant crossed field can be obtained, the probability of QED processes in such a field can be derived \cite{Ritus, RitusCC} including nonlinear Compton scattering \cite{Marklund2018}, BW pair production \cite{cascadeFedotov}, trident pair production \cite{Bentrident2013, Bentrident2018}, and so on.  These reaction data have been used extensively in the QED-PIC code for arbitrary laser configurations, based on the application of LCFA \cite{Kirk2011, Ridgers2012, Ridgers2014,Shengnc2016}. Monte Carlo simulations \cite{cascadeFedotov, cascadePiazza} show that in case a seed particle gets into a strong field with $\chi\gtrsim 1$, it can trigger a self-sustained QED cascade and the pair number grows exponentially, since the produced particles in each generation would be accelerated in the strong field and remain the same parameters on average as those of the initial seed particle.  In \cite{depletion1} it shows that the QED cascade leads to depletion of the external field, and the practically attainable strongest optical laser could be orders lower than the Schwinger critical field.

To benchmark the semiclassical approach with LCFA to the strong-field QED,  in \cite{Marklund2018} the semiclassical calculation is compared with exact results of nonlinear Compton scattering of electrons in an intense plane wave and certain agreement has been verified. As LCFA neglects the space-time variation of the external field in the formation region of the process, its validity should be scrutinized in the infrared regime where a low energy particle is involved and thus there is a possibility of a large formation length. For the nonlinear Compton scattering, it is found that the discrepancy between the accurate calculation and the calculation with LCFA increases as the emitted photon energy tends to zero and improved approximations for the photon emission probability are derived to be used in the numerical codes \cite{PiazzacorLCFA, IldertonLCFA}. The validity of LCFA is also studied for the nonlinear BW process in a strong short laser pulse  \cite{MeurenSemiclassical2016}, where an agreement with the exact calculation is found in the total probability but the main difference appears in the substructure of the spectrum, which results from interference effects between macroscopically separated formation regions.

Radiative corrections in a strong constant crossed field is found to grow surprisingly fast with $\chi$ or $\kappa$, known as Ritus-Narozhny Conjecture \cite{RitusRC1,RitusRC2, NarozhnyRC1, NarozhnyRC2}. The leading order contributions to the mass and polarization operators in such a field is $M^{(2)}(\chi\gg 1)\simeq \alpha m \chi^{2/3}$ and $P^{(2)}(\kappa\gg 1)\simeq \alpha m^2 \kappa^{2/3}$. When $\chi,\kappa \gtrsim \alpha^{-3/2}\simeq 1.6\times 10^3$, the loop corrections are not small anymore, thus making the strong constant-crossed-field QED a truly nonperturbative theory like QCD where the Furry method fails. An experiment is proposed to test the Ritus-Narozhny conjecture at $\chi\simeq 1.6\times 10^3$ by colliding a $\sim$100GeV electron with an optical laser of $I\simeq 10^{24}$W/cm$^{2}$ \cite{experimentBN}. However, calculating the mass and polarization operators in a general plane wave instead of in the constant-crossed-field approximation, it is found that the radiative corrections in a general plane wave depend algorithmically on the energy scale as in the vacuum \cite{Piazzahighenergy, Ildertonhighenergy} and thus the LCFA is not valid in obtaining the radiative correction in the high-energy asymptotic limit. It is proposed that the power-law increase of the effective coupling constant could be tested though with special experimental design to have a strong field with $\xi\gg1$ and $\chi\gg 1$ for a seed electron ($\kappa \gg 1$ for a seed photon) but with $\eta_0=k_0\cdot p/m^2\ll 1$ ($\theta_0=k_0\cdot k/m^2 \ll 1$), where $k_0$ is the four-momentum of the laser photon, $p$ is the momentum of the seed electron, and $k$ is the momentum of the seed photon \cite{Piazzahighenergy}. For example, to have $\chi=1.6\times10^3$ and $\eta=0.1$, there is $\xi=\chi/\eta\approx10^4$ which corresponds to an optical laser intensity $\simeq10^{27}$W/cm$^2$. But to keep $\eta$ small in experiments is a challenge considering the intense laser acceleration. A possible way to mitigate the influence of laser acceleration is to have a high energetic electron propagating in the same direction of $\vec{k}_0$, but then $\eta\approx \omega/(2E_p)\lesssim 10^{-6}$ for an optical laser and thus to get $\chi>1$ the required laser intensity would be unrealistically higher than the critical field. It seems that the nature makes it not easy to probe the nonperturbative loops.


\section{III. Vacuum pair production without a seed}
The proper time method invented by Schwinger is well known to beautifully obtain the analytical expression for the pair production rate in a spatially-homogeneous constant electric field, and this problem has been revisited again and again as a test ground of new developed methods. Generation of Schwinger-type pair production to more field configurations is intensively studied, such as the spatially-homogeneous temporally-oscillating electric field (OEF) \cite{Piazzareview,PhysRevA.81.022122,PhysRevLett.112.050402}. OEF can be realized around the electric anti-node of a standing wave composed of two head-on colliding laser waves of the same frequency and intensity. This is a dipole approximation reasonable only if the laser wavelength is much larger than the characteristic length scale of the pair production process, which requires $\omega\ll m$ and $\xi\gg 1$ \cite{PhysRevLett.102.080402}. This problem can be solved by the quantum kinetic method \cite{PhysRevD.90.113004}. In \cite{cascadeKim, 2016Sliva1, 2016Sliver2} the pair growth rate in a counter-propagating/multiple laser pulses is approximated by that obtained in an OEF to study the QED cascade and laser absorption in the laser-plasma interaction. A more powerful method is the Dirac-Heisenberg-Wiger formalism (DHW), which can in principle treats spatially-inhomogeneous electric field, though in practice the inhomogeneous problem is very computational costly. Detailed discussion on the theoretical methods for vacuum pair production in external electric fields is given in \cite{Gelisreview2016} and \cite{Xiereview}.

Generally speaking, there is no bottleneck problem for calculating the vacuum pair production compared to the strong dependence on the Volkov state in the Furry-Feynman method for the seed pair production problem. The theoretical simplicity is due to that photon production/absorption can be neglected here, though in the above case (e.g. see Fig. \ref{trident}$\sim$\ref{tridenttwostep}) the photon emitted by the seed particle plays an important role. Therefore, beyond a strong classical background approximation, there is no need to take into account photon quantization of other modes, and the field equation of the particle annihilation and creation operator in a classical background field can be solved by essentially solving the relativistic quantum mechanical equation, such as the Klein-Gordon equation for spin-0 particle and the Dirac equation for spin-$1/2$ particle. Pair production in an external field can be expressed as the deviation of vacuum consistency $\langle \Omega_i|\Omega_f\rangle$ from the unity. One way to compute the vacuum consistency is the world-line instaton method based on the Feynman path integral \cite{Schneider2016, Gelisreview2016}. Another way is the Bogoliubov transformation of the operators \cite{Gelisreview2016}, known also as the quantum electrodynamics with background fields \cite{unstable} or computational QED \cite{PhysRevA.97.022515}, based on the canonical operator formalism. Take electron-positron pair production for example, the main idea is: the particle creation/annihilation operators at the initial moment $t_{in}$ when the external field is not turned on yet and the final moment $t_{out}$ when the external field is turned off already are related by \cite{unstable, PhysRevD.91.125026}
\begin{equation}\label{creation_operator}
  \hat{a}_n(t_{out})=\sum_m G({}^+|{}_+)_{n;m'} \hat{a}_{m'}(t_{in})+G({}^+|{}_-)_{n;m}\hat{b}_m^\dag(t_{in}),
\end{equation}
where $\hat{a}_n(t_{out})$ is the electron annihilation operator of state $n$ at $t_{out}$, $\hat{a}_{m'}(t_{in})$ and $\hat{b}_m^\dag(t_{in})$ are respectively the electron annihilation operator of state $m'$/positron creation operator of state $m$ at $t_{in}$. The number of produced electron in state $n$ is given by
\begin{equation}\label{Num}
  \mathcal{N}_n=\left|\langle0,t_{in}|\hat{a}_n^\dag(t_{out})\hat{a}_n(t_{out})|0,t_{in}\rangle
\right|^2=\sum_{m}\left|G({}^+|{}_-)_{n;m}\right|^2,
\end{equation}
with $|0,t_{in}\rangle$ the vacuum state at the $t_{in}$ moment. In the hole theory, $G({} ^+|{}_-)_{n;m}$ is equivalent to the external-field-induced transition amplitude from the negative-energy in-state where the hole is (positron state $m$)  to the positive-energy out-state $n$. Therefore, to get the average number of pairs in each state one has to solve the time-dependent Dirac equation independently for each negative-energy state as the initial state. This method is justified when the number of created pairs is small and the interaction between the particles as well as the feedback of the particles to the laser field
can be neglected.

This method has been used to study pair production in colliding laser waves (note that a single plane-wave laser can not produce the pair). Calculations for pair production in high frequency standing waves composed of intense x-rays or gamma rays have shown drastic differences compared to OEF in the mentum spectrum and the production yield, as well as the dependence of the pair production on the laser polarization \cite{PhysRevA.97.022515,PhysRevD.96.076006,PhysRevD.97.116001,PhysRevD.91.125026,PhysRevLett.102.080402}. Based on the realization of the complete three-dimensional computation, we illustrate the coupling between the Kapitza-Dirac(KD) scattering channel and the pair creation channel in this process, and manifest the multi-channel interference effects, e.g., that similar to EIT (Electromagnetically Induced Transparency) which can lead to notable pair suppression, along with the strong field effects such as the dressed mass and nonperturbative production rates \cite{PengHu2019}.

The common limitation of the methods listed in this section is that only the external field but no quantized photons of other modes are taken into account, and thus they can not describe QED photon emission/absorption processes or compute the radiation correction. This is proposed to be improved by the real-time lattice technique \cite{Berges2011, Saffin2012, FH2013,FH2014}. Lattice QED in principle contains all QED processes in the computation. However it requires a huge amount of computing power. In \cite{FH2013} the calculation is simplified by treating the electromagnetic field in a classical-statistical way, and the evolution equations of the equal-time statistical propagator of the fermion sector are integrated stochastically using the `low-cost fermions' method \cite{Borsanyi2009}. Electron-positron pair production with the initial condition of a pure classical electric field is calculated and notable back-reaction of the created fermion pairs on the electromagnetic field is found in the critical intensity regime \cite{FH2013, FH2014}. Due to the classical statistical treatment of the gauge sector there are challenges for the simplified lattice technique to capture QED processes which produce photons of very different energies compared to the laser photon energy.

\section{IV. Summary and outlook}
The fermion pair production in a strong electromagnetic field is generally a nonperturbative QED problem. Theoretical difficulty lies not only in that the coupling coefficient gets large $\sqrt{\alpha}\sim\xi$ but also in that the effective coupling coefficient is not a constant among different photon-order expansion terms of the evolution operator. For the pair production in the combination of an energetic seed particle and a strong laser field, the ideas, applications and bottleneck problems of the Furry picture method in combination with the Volkov state are emphasized. The pole problem in computing the trident process, the strong field nonperturbative features of pair production, the generalization of the Volkov states in non-plane waves, the justification of the LCFA approximation, the vacuum cascade, and the Ritus-Narozhny conjecture problem are discussed. In view of the strong laser facilities already available, under construction, and in plan, the seed pair production has been and will continue to be the experimental realization of the pair production in the laser. Since there remain particles in the experimental vacuum, it is possible for the strong laser to accelerate the particles and trigger the vacuum cascade, restraining the attainable highest intensity of the laser to be orders lower than the critical intensity. For the vacuum pair production in the strong laser field without a seed particle, the computational QED method is emphasized and the development of the simplified lattice QED method is discussed. Pair production in the strong laser as well as other nonperturbative QED processes is an active research field both theoretically and experimentally. Efforts should continue to be devoted to derive accurate reaction probabilities (rates) required for different field scenarios. In view of the QED-PIC scheme, quantum effects and interference effects should continue to be scrutinized. Record-breaking high-energy-intensity electron-positron-photon plasma can be produced via the strong laser, which among other applications will help us understand the violent events in the universe which were inaccessible to us before.



\begin{acknowledgments}
H.H. acknowledges the support by the National Natural Science Foundation of China under Grant No. 11774415.
\end{acknowledgments}

\end{document}